\documentclass[12pt]{emulateapj}

\newcommand{\msun}{M$_{\odot}$}

\begin{document} 

\title{Blue Stragglers after the Main Sequence}
\author{Alison Sills}
\affil{Department of Physics and Astronomy, McMaster University, 1280
  Main Street West, Hamilton, ON, L8S 4M1, Canada}
\email{asills@mcmaster.ca}
\author{Amanda Karakas}
\affil{Research School of Astronomy \& Astrophysics, Mt Stromlo Observatory,
Weston Creek ACT 2611, Australia}
\email{akarakas@mso.anu.edu.au}
\author{John Lattanzio}
\affil{Centre for Stellar \& Planetary Astrophysics, Monash University,
Clayton VIC 3800, Australia}
\email{John.Lattanzio@sci.monash.edu.au}
 
\begin{abstract} 
  We study the post-main sequence evolution of products of collisions
  between main sequence stars (blue stragglers), with particular
  interest paid to the horizontal branch and asymptotic giant branch
  phases. We found that the blue straggler progeny populate the
  colour-magnitude diagram slightly blueward of the red giant branch
  and between 0.2 and 1 magnitudes brighter than the horizontal
  branch. We also found that the lifetimes of collision products on the
  horizontal branch is consistent with the numbers of so-called
  ``evolved blue straggler stars'' (E-BSS) identified by various
  authors in a number of globular clusters, and is almost independent
  of mass or initial composition profile. The observed ratio of the
  number of E-BSS to blue stragglers points to a main sequence
  lifetime for blue stragglers of approximately 1-2 Gyr on average.
\end{abstract} 

 
\section{INTRODUCTION} \label{intro} 
 
Blue stragglers are main sequence stars that are bluer and brighter
than the main sequence turnoff in clusters. They are expected to have
been formed through either dynamical processes, such as direct stellar
collisions or during the disruption of a binary system; or through the
coalescence of an unperturbed primordial binary in the cluster. 

Blue straggler populations have been used to
try to and constrain the dynamical evolution of globular clusters
\citep[e.g.][]{1993ApJ...403..256H}. Since the number of collision
products in a cluster should tell us about the rate of collisions in
the past, it follows that we should be able to constrain the density,
mass function, or other cluster properties from the blue straggler
population. In practice this has been difficult, since the number of
unsettled questions about the formation mechanisms of blue stragglers,
and competing effects of cluster evolution, is large. However,
\citet{2000ApJ...535..298S} and \citet{2006ApJ...650..195M} claim,
based on the blue straggler population, that 47 Tucanae may have come
close to core collapse a few Gyr ago; and \citet{1999ApJ...522..983F} 
use the blue straggler population of M80 to postulate imminent core 
collapse in that cluster as well. 
 
But what happens after the blue stragglers evolve off the main 
sequence? Can we see them anywhere in the colour-magnitude diagram 
(CMD), and if we can, does that give us any further information about 
the past of the cluster? Blue stragglers have lifetimes of approximately 
1 Gyr or so \citep{1997ApJ...487..290S}, and it would be 
useful to probe slightly further back in time if we can. 

There is a population of HB stars in clusters slightly brighter than 
the regular horizontal branch that has been seen in M3 as early as 
\citet{1953AJ.....58...61S}. 
Blue stragglers are expected to have  masses of up to twice the 
turnoff mass in globular clusters, and using these masses in standard
single-star evolutionary models puts them in this part of the 
HB branch of the CMD \citep{1988ARA&A..26..199R}: BSS
progeny should be redder and slightly brighter than their low-mass
counterparts. Knowing the relative numbers of blue stragglers and
their evolved progeny can constrain the lifetimes of these two stages.
\citet{1986MmSAI..57..453I} also suggests that different formation
mechanisms for blue stragglers could result in very different
properties later in their evolution, and so observing evolved blue
stragglers may help us constrain their formation.

Evolved blue stragglers should also exist in the instability strip,
and the expectation is that they could be observed as Anomalous
Cepheids. These stars are low metallicity pulsating stars
approximately 2 magnitudes brighter than RR Lyrae stars, and with
pulsation periods around 1.5 days. Almost all are found in dwarf
galaxies in the Local Group. They are expected to be the result of
mass transfer and possibly coalescence in a low mass binary system (up
to about 1.6 \msun) \citep{1976ApJ...209..734Z,1988ApJ...324.1042C},
which is exactly the same as one scenario for the formation of blue
stragglers.  Only one anomalous Cepheid has been observed in a
Galactic Globular Cluster: V19 in NGC 5466
\citep{1982ApJ...262..700Z}. A detailed spectroscopic study of that
star \citep{1997ApJ...482..203M} found a mass of 1.66 +0.7-0.5 \msun,
a small rotational velocity $v \sin i = 18$ km s$^{-1}$, normal iron,
s-process, and alpha element abundances for a cluster member, and some
evidence for a very long-period (P $\sim$ 10 000 days) binary orbit.
The rotational velocity, chemical composition, and orbital period
argue against a recent Roche Lobe overflow mass transfer episode, but
are not inconsistent with a collisional origin for the blue straggler.
Unfortunately, NGC 5466 is quite a sparse cluster and therefore
collisions are rather unlikely. Nevertheless, the existence of an
unusual evolved star as well as a significant number of blue
stragglers in this cluster makes it an interesting case.
 
The first paper to explore the possible connection between blue
stragglers and horizontal branch morphology in many clusters was
\citet{1992AJ....104.1831F}.  They looked at 21 globular clusters with
both blue stragglers and well-observed horizontal branches.  They
argue that the reddest horizontal branch stars, particularly in
clusters with bluer HBs, are the descendents of blue stragglers.
Based on their analysis of the observations of BSS and HB stars in 10
globular clusters, they determine that the average ratio of the number
of blue stragglers to the number of ``blue straggler progeny'' on the
HB to range from 1 to 10, with a mean of 6.6 and an overall value of
5.3, summed over the 10 clusters.

\citet{1994AJ....107.1073B} studied the evolved stars in the nearby globular
cluster 47 Tucanae, using a combination of ground-based and Hubble
Space Telescope (WFPC) observations to look at the distribution of
these stars throughout the cluster. He found an enhancement of
asymptotic giant branch (AGB) stars, approximately one magnitude brighter
than the horizontal branch, in the centre of this cluster only. Since
the blue stragglers in that cluster (and many others) are centrally
concentrated, he makes the link between these two populations. More
recently, \citet{2006ApJ...652L.121B} used HST/ACS to look at the
evolved stars in 47 Tucanae with better precision than the study of
\citet{1994AJ....107.1073B}. They confirm Bailyn's result of a central
enhancement of AGB stars, which they interpret as the result of
evolved blue stragglers formed in binary mergers. Neither study was
sensitive to stars much fainter than the HB, and so do not give the
number of blue stragglers, or the ratio of blue stragglers to HB or
AGB stars. 
 
Blue stragglers and their evolved counterparts (E-BSS) have been
identified using HST/WFPC2 in M3 \citep{1997A&A...320..757F}, M13
\citep{1997ApJ...484L.145F, 1999ApJ...522..983F} and M80
\citep{1999ApJ...522..983F}. The E-BSS are identified as being
slightly bluer than the RGB and between 0.2 and 1 magnitudes
brighter than the horizontal branch in those clusters.  In M3 and M80,
the radial distribution of the blue stragglers and E-BSS is the same,
while both populations are more centrally concentrated than the normal
RGB stars in those clusters. The ratios of `bright' blue stragglers to
E-BSS are determined from these studies to be 6.4 in M3, 2.2 in M13,
and 6.8 in M80 In M80, the population of blue stragglers is so well
determined that the limitation to bright blue stragglers is not
necessary. In that case, the ratio of all blue stragglers to E-BSS is
$\sim$ 16.

In this paper, we use models of stellar collision products as starting
models for stellar evolutionary calculations. We follow the evolution
of these stars from the main sequence to the giant branch, through the
helium flash onto the horizontal branch, and onto the asymptotic giant
branch. We look at their positions in the HR diagram and their
lifetimes, and compare to the observational evidence listed above. We
wish to determine whether the stars identified in the observational
studies are indeed evolved blue stragglers, and whether their observed
properties agree with our models of blue straggler formation and
evolution. We outline our computational method in section
\ref{method}, and present the results in section \ref{results}. The
implications and a discussion of further questions follows in
section \ref{discussion}.

\section{METHOD} \label{method}

In this paper, we assume that blue stragglers were formed during
direct collisions between two unrelated main sequence stars. The other
methods of forming blue stragglers (collisions mediated by binary
stars, and mass transfer in binary systems) will probably result in
somewhat different structure and composition profiles of the
newly-born blue stragglers \citep[see e.g.][]{1995ApJ...439..705B}.
However, the direct collision products are the best-studied blue
stragglers to date
\citep{1997ApJ...487..290S,1997ApJ...477..335S,2001ApJ...548..323S},
and also the easiest to model. The possible implications of this
choice on our results will be discussed in section \ref{discussion}.

\subsection{The collision parents}

We used the Yale Rotational Evolutionary Code
\citep[YREC,][]{1992ApJ...387..372G} to calculate evolutionary tracks for
representative globular cluster stars. We chose a metallicity of
Z=0.001 and Y=0.232 (corresponding to [Fe/H]=-1.27 where Z$_{\odot}$ =
0.0188). Our mixing length is $\alpha=1.642$, calibrated to give the
solar luminosity and radius at the solar age for a solar-mass star
with solar metallicity.  We ran models for 0.4 \msun, 0.6 \msun and
0.8 \msun~ stars, starting at the deuterium-burning birthline
\citep{1991ApJ...375..288P}, and followed the evolution through the
pre-main sequence and main sequence phases.

We are interested in collision products which would show up in the
blue straggler region of the colour-magnitude diagram at the current
time, or in the recent past. We assume that the typical mass of
turnoff stars in globular clusters is 0.8 \msun. From our evolutionary
models, the age of a 0.8 \msun~ star at hydrogen core exhaustion is
13.7 Gyr, so we choose this value to be a typical age of a globular
cluster.

In previous work \citep[e.g.][]{1997ApJ...487..290S}, the parent stars were
assumed to have the current age of the cluster. In other words, all
collisional blue stragglers were formed from parents whose composition
profile was that of a 13.7-Gyr-old star. However, collision products
can live on the main sequence for quite a long time -- up to an
additional 10 Gyr for a low-mass collision product of 0.8 or 0.9
\msun. Therefore, we should be including collision products whose
parents collided quite some time ago, and whose structure was that of
a much younger star. The thermodynamic and physical structure does not
change much on the main sequence, but the chemical composition profile
of the star is modified by nuclear burning.  For the 0.4 \msun~ models,
this effect is not very strong since they have completed only a small
fraction of their main sequence lifetime. However, the 0.8 \msun~
models have significantly different amounts of helium in their cores
if we look at the models at an age of 10 Gyr compared to 13.7 Gyr. The
younger star has Y=0.855 at its centre, compared to Y=0.998 in the
older turnoff star.

\begin{figure} 
\includegraphics[width=\columnwidth]{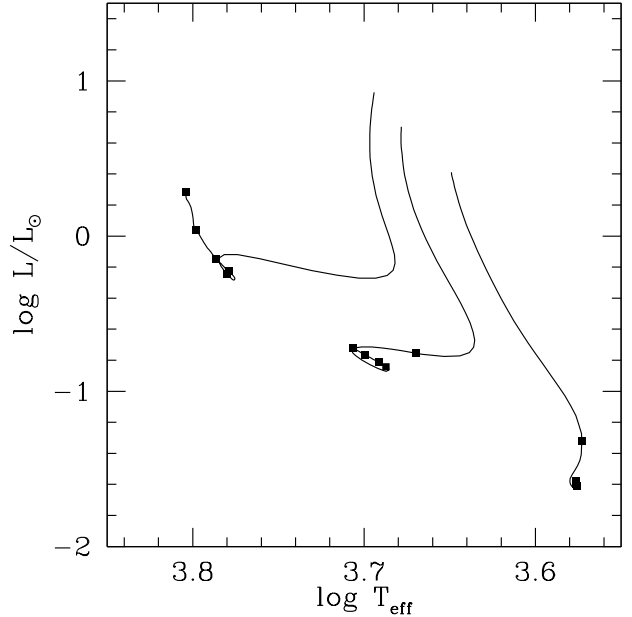} 
\caption{Evolutionary tracks of the parent stars of the collision 
  products: 0.4, 0.6 and 0.8 \msun. The stars begin their lives on the
  deuterium burning birthline, contract towards the main sequence, and
  then remain on the main sequence for the full 13.7 Gyr of evolution.
  The properties of the stars at the possible times of collisions are
  marked with large squares, with points A through E moving from the
  pre-main sequence along the main sequence evolution.}
\label{fig:parenttracks} 
\end{figure}

We chose 5 different ages of parents to investigate this effect: 0.03
Gyr (corresponding to the amount of time it takes a 1.0 \msun~ star to
reach the zero-age main sequence from the birthline), 2 Gyr, 5 Gyr, 10
Gyr and 13.7 Gyr. We labeled these ages as A, B, C, D and E when
discussing our collision products in the rest of the paper. The
evolutionary tracks, and the positions of the parents at these ages,
are show in figure \ref{fig:parenttracks}.

In this paper, we looked at all collision products which could
plausibly be seen in the colour-magnitude diagram of a globular
cluster -- those that could be on the main sequence, giant branch,
horizontal branch or asymptotic giant branch at the current time. To
determine which collision products to concentrate on, we estimated the
main sequence lifetime of each product after the collision. If the sum
of the time of the collision plus the main sequence lifetime of the
product was 10 Gyr or longer, we included those stars in our
investigation. For example, a collision between two 0.8 \msun~ stars at
time B (2 Gyr after the cluster was formed) results in a collision
product which is approximately 1.5 to 1.6 \msun. The main sequence
lifetime of a 1.5 \msun~ star is something like 4 Gyr, and so this
particular collision product will not be around at the present time.
In fact, the only collisions between two 0.8 \msun~ stars that could
possibly be observed now are those which occurred at times D and E in
this scheme. Collisions between two 0.4 \msun~ stars, on the other
hand, produce stars with masses of 0.8 \msun~ or less, and all of those
collision products should still be on or close to the main sequence.
 
The collisions that we will discuss in this paper are given in Table
\ref{table:collisions}. We give the name of each collision, the masses
of the two parent stars, the age of the parents at the time of
collision, and the mass of the collision product. 

\begin{deluxetable}{lcccccccc} 
\tablecaption{Initial Parameters of Collision Products \label{table:collisions}} 
\tablehead{ 
\colhead{Collision Name} & \colhead{M$_1$} & \colhead{M$_2$} & 
\colhead{Time of Collision} & \colhead{M$_{\rm tot}$}\\ 
\colhead {} & \colhead{\msun} & \colhead{\msun} & \colhead{Gyr} & 
\colhead{\msun}  
} 
\startdata 
m04m04A & 0.4 & 0.4 & 0.03 & 0.78 \\ 
m04m04B & 0.4 & 0.4 & 2    & 0.78 \\ 
m04m04C & 0.4 & 0.4 & 5    & 0.78 \\ 
m04m04D & 0.4 & 0.4 & 10   & 0.78 \\ 
m04m04E & 0.4 & 0.4 & 13.7 & 0.78 \\
m04m06C & 0.4 & 0.6 & 5    & 0.98 \\
m04m06D & 0.4 & 0.6 & 10   & 0.98 \\
m04m06E & 0.4 & 0.6 & 13.7 & 0.98 \\
m04m08D & 0.4 & 0.8 & 10   & 1.18 \\
m04m08E & 0.4 & 0.8 & 13.7 & 1.18 \\
m06m06D & 0.6 & 0.6 & 10   & 1.17 \\
m06m06E & 0.6 & 0.6 & 13.7 & 1.18 \\
m06m08D & 0.6 & 0.8 & 10   & 1.37 \\
m06m08E & 0.6 & 0.8 & 13.7 & 1.38 \\
m08m08D & 0.8 & 0.8 & 10   & 1.57 \\
m08m08E & 0.8 & 0.8 & 13.7 & 1.58 \\
\enddata
\end{deluxetable}
 
\subsection{The collision product and evolution to the Main Sequence}

We used the code Make Me A Star \citep[MMAS][]{2002ApJ...568..939L} to
calculate the structure and chemical profile of our collision
products. MMAS takes the results of stellar evolution calculations
(from YREC or other codes), and uses a method called ``sort by
entropy'' to determine the structure and chemical profile of the
collision product. It has been shown that the results of low-velocity
stellar collisions can be readily approximated by determining the 
entropy of each shell in the parent star (using $A=P/\rho^{\Gamma}$) 
and then creating the collision product by sorting the shells of both 
parents stars so that entropy increases outward. This technique
assumes that the collision does not shock-heat much of the material
(thereby changing its entropy), and was based on the results of smoothed
particle hydrodynamic simulations of stellar collisions. 

We assumed that all collisions occurred with a periastron separation of
0.5 times the sum of the radii of the two stars. It has been shown
that the mass and structure of the collision products does not depend
strongly on the impact parameter, except that head-on collisions are
different than off-axis collisions
\citep{1996ApJ...468..797L,1997ApJ...487..290S}.  Since head-on
collisions are very rare, we chose to use a representative off-axis
collision.

We used the method outlined in \citet{1997ApJ...487..290S} to take the
collisions products and evolve them from the end of the collision to
the main sequence. For this portion of the evolution, we used YREC. In
this stage, and throughout the paper, we neglect any rotation of the
collision product in our evolution. We stopped the evolution when the
energy generation due to hydrogen burning was larger than that due to
gravitational contraction, and the star had reached the faintest
luminosity in that region of the CMD. This is equivalent to the
definition of the Zero Age Main Sequence for a normal pre-main
sequence evolutionary track. This post-collision stage lasted between
$10^7$ and $10^8$ years, depending on the mass of the collision
product.

\subsection{Evolution from the Main Sequence} 

We wish to follow these collision products on the main sequence and
giant branch, onto the horizontal branch and then to the asymptotic
giant branch. These products are low mass stars, and so they will
undergo the helium flash at the tip of the giant branch. YREC, like
most stellar evolution codes, cannot follow the evolution of a star
through the helium flash because of the very short timescales
involved and large helium burning luminosities, up to $10^{10} L_{\odot}$,
depending on the initial mass and $Z$.  

The Monash stellar evolution code, however, can evolve stars through
the core helium flash
\citep[e.g.,][]{karakas02}\footnote{\citet{lattanzio86} created
  zero-age horizontal branch models instead of evolving through the
  flash.}.  This is because the code has been specifically adapted
\citep{lattanzio86,lattanzio91,frost96} to model the evolution of AGB
stars and the conditions during instabilities of the He-burning shell
\citep{schwarzschild65} are similar, though not as extreme, as found
in the cores of low-mass stars during He ignition.


Neutrinos losses are also an important component when modeling 
the core helium flash because they move the temperature maximum, 
and hence the ignition point, outward from the stellar center. 
Neutrinos losses included in the Monash code are  
described in detail in \citet{lattanzio86}. Note that the maximum 
stellar mass that experiences the core helium flash decreases
as a function of metallicity; this is because these models 
are hotter owing to a lower opacity; at $Z=0.001$ this 
mass is $\sim 2$ \msun~ compared to 2.25~\msun~ at $Z=0.02$. 
We stress that although we can use the Monash evolution 
code to evolve models through the core helium flash, the  
models should only be considered as crude approximations, 
given that the details of the evolution depend critically  
on the assumptions made about convective energy transport,
that is clearly not one dimensional in nature. 
A thorough understanding about this complicated phase 
will require detailed multi-dimensional hydrodynamical
modeling, see \citet{dearborn06} for recent efforts.
 
We converted the results of the YREC calculations into the format
required by the Monash code. We assumed that the star
was initially fully convective.  The Monash code determines which
shells of the star are convective before doing any subsequent mixing
or evolution, so this assumption was purely for simplicity. To
calculate the internal energy, we assumed that the equation of state
was that of an ideal monatomic gas plus radiation pressure, which is a
good approximation to main sequence low mass stars. The mass, radius,
luminosity, pressure, temperature, density and compositions of H,
$^3$He, $^4$He, C, N, O and Z were taken from the YREC models
directly. These new models were then used as starting models, and the
collision products were evolved from the main sequence until the
asymptotic giant branch. In some cases, the evolution continued
through the thermal pulse phase and onto the post-AGB and white-dwarf
cooling track.

In the Monash evolution code we included mass loss using the 
\citet{reimers75} formula   
\begin{equation} 
  \dot{M} = - 4 \times 10^{-13} \eta \frac{L R}{M} \,\,\, M_{\odot}\,{\rm year}^{-1} 
\end{equation}  
where $L, R$ and $M$ are the stellar luminosity, radius and mass,  
respectively, and the parameter $\eta$ varies between 0.4 $< \eta$ 4,  
although values as high as 10 have been used \citep{straniero97}.  
We set the parameter $\eta = 0.4$ on the RGB and $\eta =1$ on the AGB.  
To test the effect of mass loss on the AGB lifetime in the  
collision-product runs, we also used the \cite{vw93} formulation on the  
AGB, that relates the fundamental pulsation period of the  
star to the mass-loss rate, and includes a superwind phase once 
the period increases above 500~days. We assumed instantaneous 
mixing, and used the same mixing length as used in YREC, $\alpha = 1.642$. 
For more details of the input physics used in the Monash evolution 
code we refer the reader to \citet{lattanzio86,frost96} and 
and \citet{karakas07b}. 

\section{Results} \label{results}

\begin{figure}
\includegraphics[width=\columnwidth]{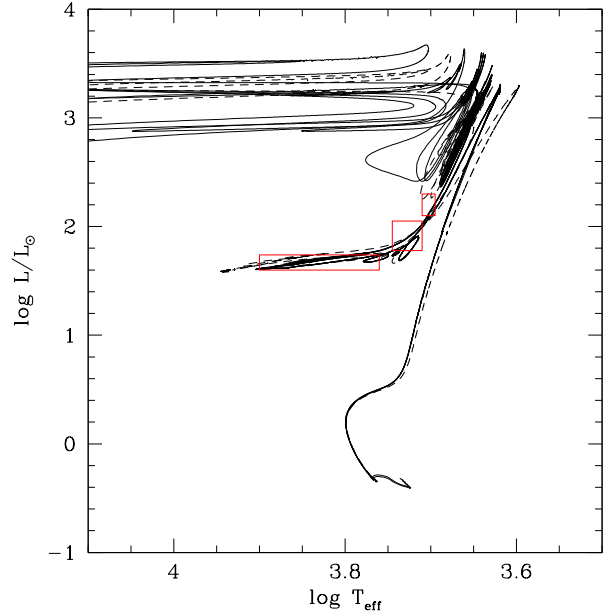}
\caption{Evolutionary tracks of all the m04m04 stellar collision
  products discussed in this paper, plus the m=0.78 \msun normal star
  track (marked with a dashed line). The other lines are the
  different versions of the same parent stars, collided at different
  times in the parents' lives. }
\label{fig:m04m04tracks}
\end{figure}

\begin{figure} 
\includegraphics[width=\columnwidth]{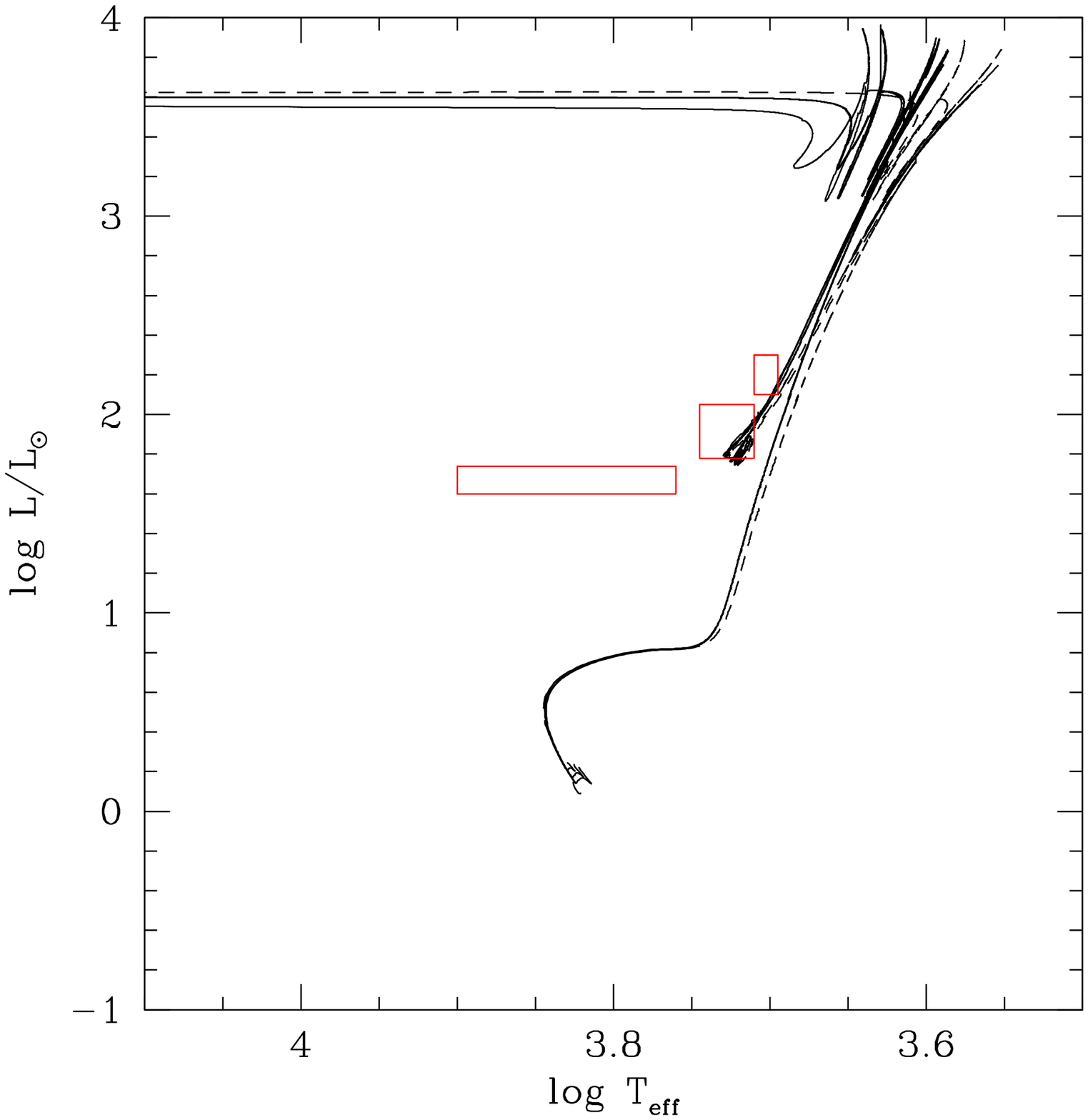} 
\caption{Evolutionary tracks of all the m04m06 stellar collision products 
  discussed in this paper, plus the m=0.98 \msun normal star. The lines are the same as in figure \ref{fig:m04m04tracks}.} 
\label{fig:m04m06tracks} 
\end{figure} 

\begin{figure}
\includegraphics[width=\columnwidth]{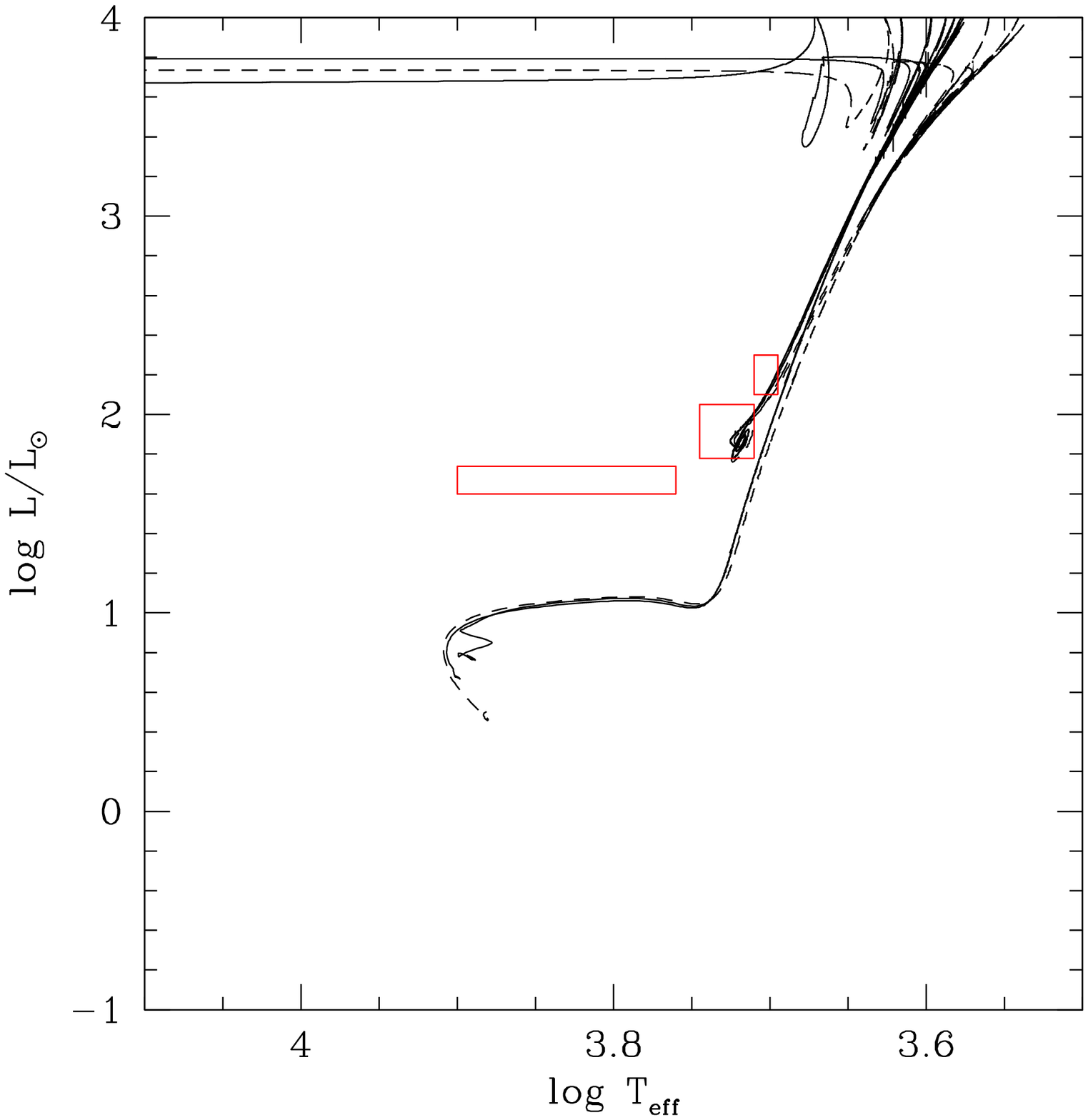}
\caption{Evolutionary tracks of all the m04m08 stellar collision products
  discussed in this paper plus the m=1.18 \msun normal star. The lines are the same as in figure \ref{fig:m04m04tracks}.}
\label{fig:m04m08tracks}
\end{figure}

\begin{figure}
\includegraphics[width=\columnwidth]{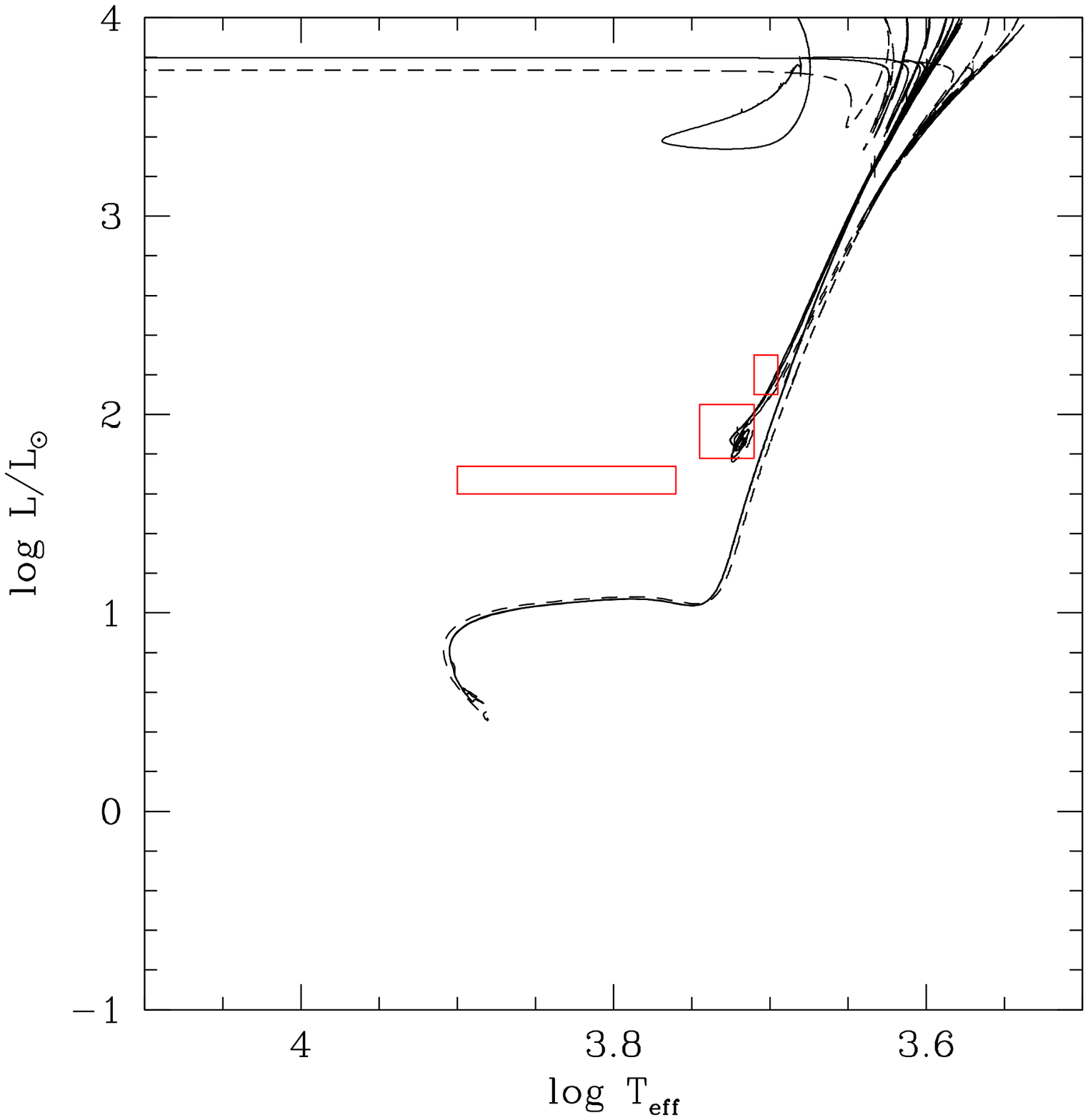}
\caption{Evolutionary tracks of all the m06m06 stellar collision products
  discussed in this paper plus the m=1.18 \msun normal star. The lines are the same as in figure \ref{fig:m04m04tracks}.}
\label{fig:m06m06tracks}
\end{figure}

\begin{figure} 
\includegraphics[width=\columnwidth]{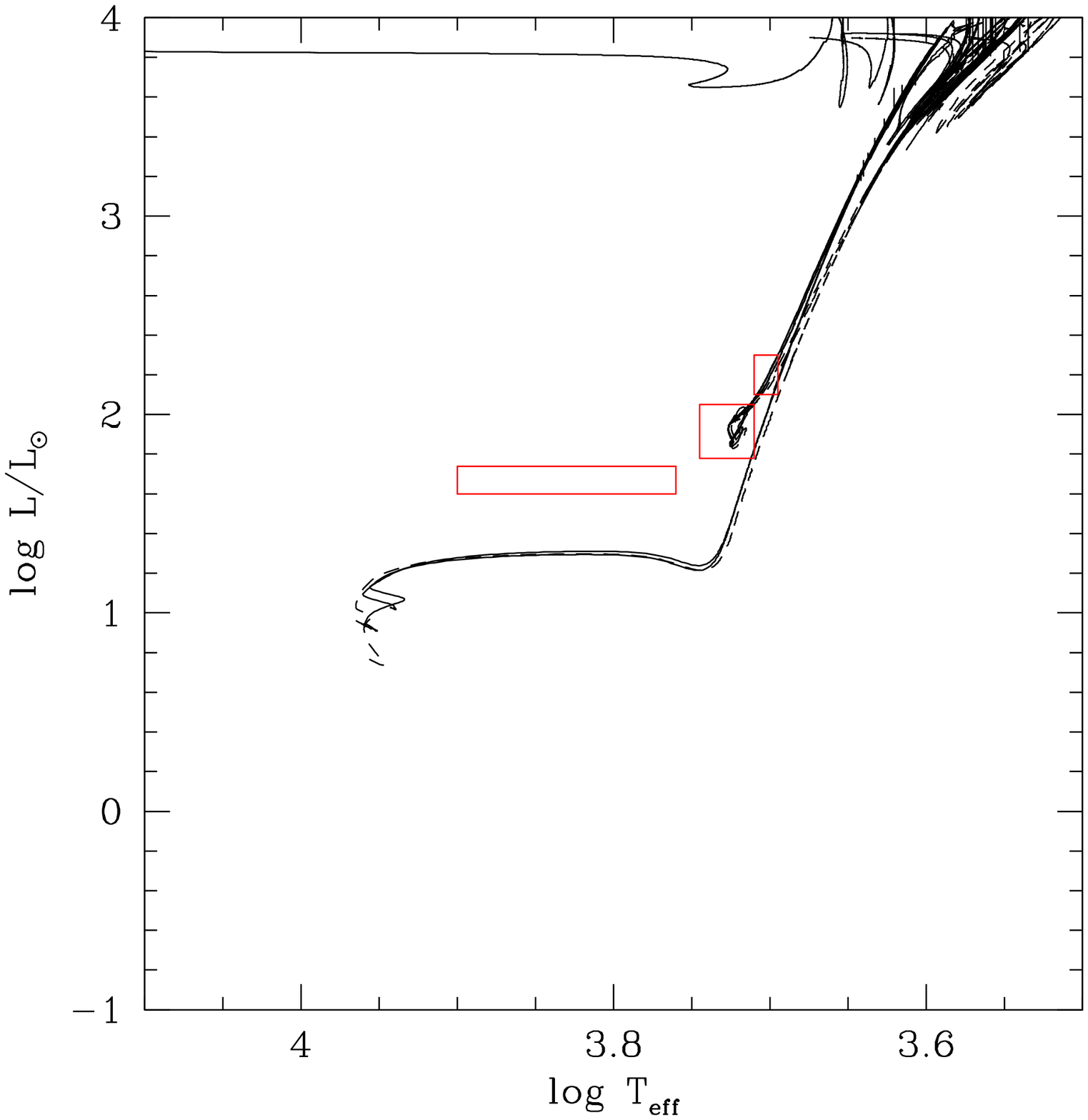}
\caption{Evolutionary tracks of all the m06m08 stellar collision products
  discussed in this paper plus the m=1.38 \msun normal star. The lines are the same as in figure \ref{fig:m04m04tracks}.}
\label{fig:m06m08tracks}
\end{figure} 

\begin{figure}
\includegraphics[width=\columnwidth]{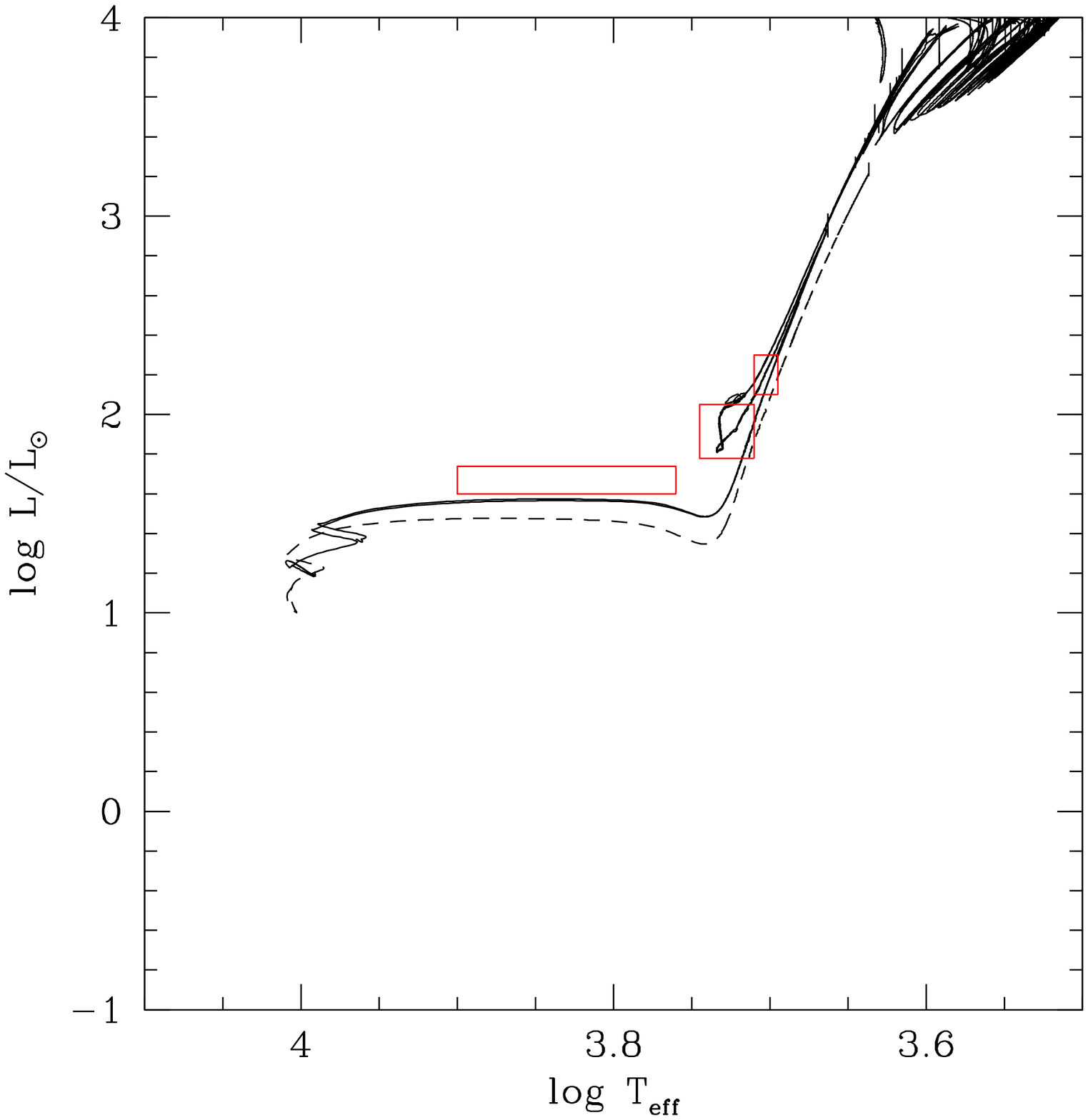}
\caption{Evolutionary tracks of all the m08m08 stellar collision products
  discussed in this paper plus the m=1.58 \msun normal star. The lines are the same as in figure \ref{fig:m04m04tracks}.}
\label{fig:m08m08tracks}
\end{figure}

\subsection{Comparison to Normal Stars}

In figures \ref{fig:m04m04tracks} -- \ref{fig:m08m08tracks}, we
present evolutionary tracks of each of our collision products.
Each figure corresponds to a different choice of parent star masses. 
The different lines correspond to the different ages at which
the parent stars collided.  We also present evolutionary tracks of
normal stars of similar masses to our collision products, shown in the
figures as dashed lines. The normal stars have a uniform chemical
composition throughout the star at the Zero Age Main Sequence, with
$Z=$0.001 and $Y=$0.232, the same as our parent stars.

In all these figures, we have marked the approximate position of the
observed ``E-BSS'' stars, as well as the horizontal branch and the
base of the asymptotic giant branch (where these stars are most likely
to be found observationally). The HB was defined by the evolutionary
track of the m=0.78 \msun track, and lies at a luminosity of $\log
(L/L_{\odot}) \sim 1.7$.  The AGB box was chosen to lie just above the
start of the AGB (as defined below ) for most of our tracks.

The E-BSS stars are, observationally, found somewhere between 0.2 and
1 magnitudes brighter than the horizontal branch
\citep{1994AJ....107.1073B, 1997A&A...320..757F,1999ApJ...522..983F},
and just to the blue of the RGB. Because we are working in the
theoretical plane, we need to convert from magnitude differences to
luminosity differences: one magnitude is equal to a difference of
0.4 in $\log L$. Therefore, our E-BSS box goes from $\log 
(L/L_{\odot}) = 1.78$ to $2.1$, and we chose the temperature extent
to begin just blueward of the RGB and extend for a few hundred
degrees.

The main sequence position of each track in the CMD is very similar in 
each figure. Choosing a different time of collision for the parent 
stars makes a small difference to the position of the collision 
product in the CMD. The evolutionary tracks are also very similar to 
their normal main sequence counterpart. As has been shown in previous  
work \citep{1997ApJ...487..290S}, the 
collisions involving lower-mass parents show less effect on the main 
sequence. The lower-mass parents are less evolved overall. The 0.8 
\msun~ stars are at the turnoff at time `E', whereas the 0.4 \msun~ 
stars are simply a few more Gyrs along in their $\sim$ 40 Gyr 
evolution. 
 
The post-main sequence evolution of the normal stars and the collision 
products is very similar. There is some slight colour difference 
between the normal stars and collision products on the giant branch, 
with the normal stars being slightly redder. The difference is usually 
a few tens of degrees at a given luminosity, and is probably not 
observable. The largest different is $\sim$ 300 degrees at the tip of 
the giant branch in figure \ref{fig:m04m04tracks}, and even this will 
be difficult to observe since there should be essentially no blue 
stragglers in that evolutionary state in most clusters.

The location of the core-helium-burning phase for higher-mass stars is
consistent with the nominal observed position of the E-BSS stars in
the CMD. For all the collision products except the m04m04 series, the
E-BSS box covers the majority of the track during core-helium-burning.

\subsubsection{Horizontal Branch Evolution}

\begin{figure}
\begin{tabular}{c}
\includegraphics[angle=270,width=\columnwidth]{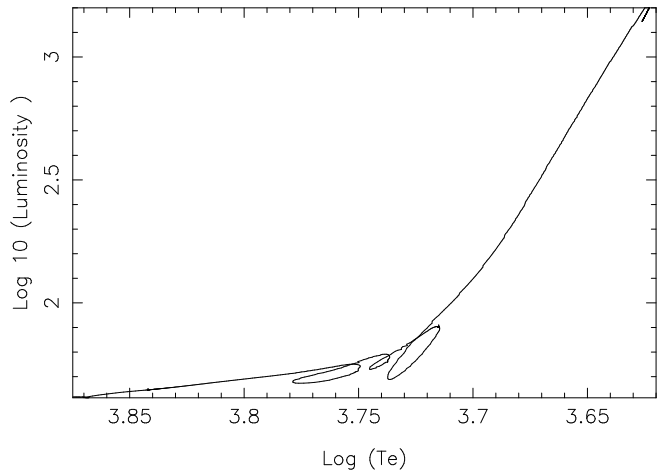} 
\end{tabular}
\caption{Evolutionary track of the m04m04B model from 
the tip of the giant branch to the start of quiescent
core helium burning.}
\label{fig:loops}
\end{figure}

In the CMDs for both normal stars and collision products, we see some
loops in the CMD on the early part of the horizontal branch, 
see figure~\ref{fig:loops}.  These loops are a product of taking
the star through the helium flash, rather than stopping the evolution
at the tip of the giant branch and restarting it with a zero-age 
horizontal branch model, as is done by a
number of groups \citep{charbonnel96,castellani92,stancliffe05a}. 
The timescale for these loops is very short ($< 10^{6}$ years), 
and they occur at the end of the helium flash but before stable
core helium burning has begun.

\begin{figure}
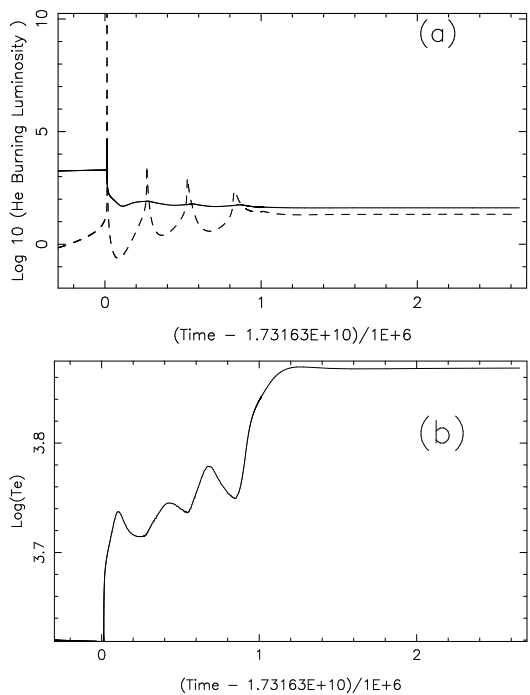

\begin{tabular}{c}
\includegraphics[angle=270,width=.8\columnwidth]{f9a.eps} \\
\includegraphics[angle=270,width=.8\columnwidth]{f9b.eps} 
\end{tabular}
\caption{(A) Surface (black) and helium-burning (red) 
luminosity as a function of scaled time 
for the m04m04B model from the tip of the
giant branch to the start of quiescent
core helium burning.  Time is scaled such that the 
beginning of the flash is $t = 0$, with time in units
of $10^{6}$ years. Most of the energy produced 
by the core helium flash (maximum of $\sim 10^{10} L_{\odot}$ 
at first flash) goes into lifting the electron degeneracy, 
with some converted  into mechanical energy to expand 
the star. (B) Logarithm of the effective temperature.
The time from the flash peak to quiescent core He-burning
is approximately $10^{6}$ years.}
\label{fig:heflash}
\end{figure}

The loops are caused by successive oscillations of the core as it
settles down after the core He-flash. In figure~\ref{fig:heflash}(a)
we show the surface and helium-burning luminosity from the m04m04B
collision model during the core helium flash.  Illustrated are
successive flashes of lessening intensity, these occur because the
core helium flash is off center to begin with, and each successive
flash moving inward (in mass) and removing degeneracy. This is similar
to the degenerate carbon ignition found in 8 to 11~\msun~ stars
\citep{siess06}.  In figure~\ref{fig:heflash}(b) we see the change to
the effective temperature from the tip of the GB to the beginning of
quiescent core helium burning. The effective temperature first
increases when the flash begins, followed by short timescale, small
amplitude oscillations caused by the whole star expanding and
contracting from the successive core flashes. Once the star begins
quiescent core helium burning the effective temperature stays
approximately constant. We do not assume any extra mixing, convective
overshoot, etc. from the border of the core during this phase.  

\subsection{Lifetimes and Possible Evolved Blue Straggler Stars}

We are interested in how many post-main sequence blue stragglers we
might expect. One easy way to approximate this is to look at the
timescales of each of the different stages in the CMD, and to equate
the relative number of objects in each stage to the ratio of the
lifetimes. First we need to clearly define each stage. We took the
``zero age'' main sequence to be the start of the calculation using
the Monash code. At this point, the stars have come into hydrostatic
equilibrium, and are burning hydrogen in their cores. We define the
end of the main sequence (Terminal Age Main Sequence) to be the age at
which the central helium mass fraction $Y$ has reached 0.99. We are also
interested in the age that defines the beginning of the horizontal
branch, which we take to be the age at which $Y$ drops to 0.97 after the
TAMS. The end of the horizontal branch phase occurs when the central $Y$
drops to zero, and we also take this to be the start of the asymptotic
giant branch phase. 

\begin{figure}
\includegraphics[width=\columnwidth]{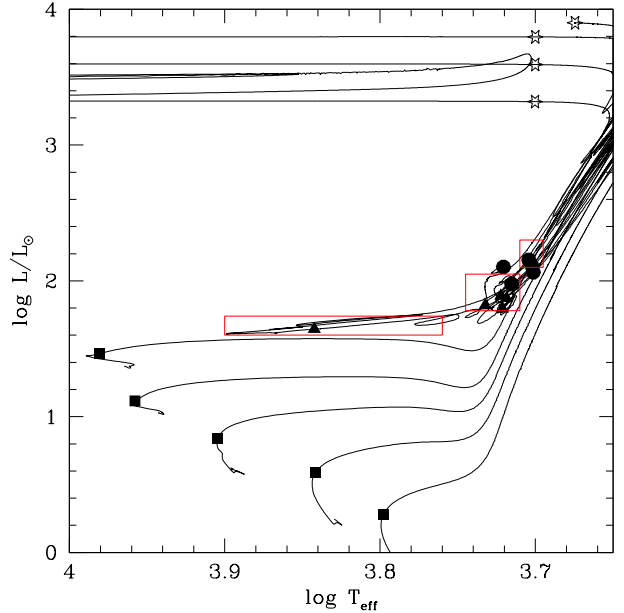}
\caption{Evolutionary tracks of a few stellar collision products. From
  faintest to brightest, they are: m04m04E, m04m06E, m06m06E, m06m08E,
m08m08E. The zero-age main sequence is taken to be the beginning of
these tracks. The terminal-age main sequence is marked with squares;
the zero-age horizontal branch is marked with triangles; and the
terminal-age horizontal branch is marked with circles. The end of the
asymptotic giant branch phase is marked with stars.}
\label{fig:sometracks}
\end{figure}

Mass loss terminates asymptotic giant branch evolution
when the envelope mass is reduced to $\approx$~0.01\msun 
\citep{bloecker01}, or when $q = M_{\rm core} / M_{\rm total} > 0.9$
with $q$ approaching unity as the star evolves to the white dwarf (WD)
cooling track \citep{schoen79}. We choose to define a star to
have left the AGB track when the effective temperature, $T_{\rm eff}$, 
increases by $\Delta \log T_{\rm eff} =0.3$ on the CMD \citep{karakas07b}.
From figure~\ref{fig:sometracks} we can approximate the location of 
the AGB to be $\log T_{\rm eff} = 3.7$. 
Some models evolved to the WD cooling track, so to
obtain an accurate estimate of the AGB lifetime, we needed to remove
the time spent during the post-AGB (usually short, 
$\lesssim 10^{5}$years) plus on the WD cooling track. For other
models, convergence difficulties ended the calculation before
all of the envelope mass was lost and the model did not move
away from the AGB. For all the collision models, however, 
$q \gtrsim 0.95$ and all final envelope masses are
few times $10^{-2}$\msun or less (maximum is 0.034\msun).
In particular, all Reimer's models had $q > 0.98$ except
the m08m08D model, that had $q = 0.976$ and an envelope
mass of 0.016\msun.
For models that had their evolution halted by convergence
difficulties, the final envelope mass was small enough that we can  
assume that no more thermal pulses would occur, and that the
final timestep is an excellent approximation to the final AGB
time. This is further justified because the thermally-pulsing
phase is itself a small fraction of the total AGB lifetime. 
For example, the total AGB lifetime of the 1.58\msun, $Z=0.001$ 
normal model is 488.6~Myr, much longer than the thermally-pulsing AGB 
lifetime of only 2.1~Myr. The final envelope mass was 
$\approx 0.03$\msun, which would be lost in much less
than the time taken between successive shell flashes. 

In figure \ref{fig:sometracks}, we show a few evolutionary tracks, 
with these important points marked with solid symbols.  The duration
of each phase for all our collision products is given in table
\ref{table:lifetimesReimers}, and for the normal stars in table
\ref{table:lifetimesNormal}. In all cases, the mass loss on the AGB
was calculated using the Reimer's formalism. The first column of each
table identifies the star or collision product. The next three columns
give the main sequence, horizontal branch, and asymptotic branch
lifetime in Gyr. The final two columns give the ratio of main sequence
to horizontal branch lifetime, and the main sequence to asymptotic
giant branch lifetime. These last two columns are an indication of the
ratio of the numbers of those kinds of stars, under the assumption of
a constant formation rate.

\begin{deluxetable} {lcccrr} 

\tablecaption{Lifetimes and Population Ratios of Evolved Collision Products \label{table:lifetimesReimers}}
\tablehead{
\colhead{Collision} & \colhead{MS} & \colhead{HB} & \colhead{AGB} & \colhead{MS/HB} & \colhead{MS/AGB} \\
 & \colhead{Gyr} & \colhead{Gyr} & \colhead {Gyr} & & \\
}
\startdata
     m04m04A &  14.1  & 0.095 &  0.017 &   148.8 &   846.3 \\
     m04m04B &  14.1  & 0.095 &  0.017 &   148.3 &   855.2 \\
     m04m04C &  14.3  & 0.097 &  0.016 &   146.6 &   906.5 \\
     m04m04D &  14.4  & 0.095 &  0.016 &   150.8 &   889.7 \\
     m04m04E &  14.0  & 0.097 &  0.016 &   144.3 &   883.8 \\
     m04m06C &  5.73  & 0.093 &  0.016 &   61.7 &   404.8 \\
     m04m06D &  5.06  & 0.094 &  0.014 &   53.6 &   343.4 \\
     m04m06E &  4.40  & 0.089 &  0.015 &   49.3 &   276.2 \\
     m04m08D &  0.93  & 0.093 &  0.016 &   10.0 &   65.6 \\
     m04m08E &  0.60  & 0.090 &  0.014 &   6.7  &   39.8 \\
     m06m06D &  2.18  & 0.092 &  0.015 &   23.7 &   146.6 \\
     m06m06E &  1.81  & 0.091 &  0.015 &   19.9 &   120.5 \\
     m06m08D &  0.82  & 0.096 &  0.014 &   8.5 &   56.8 \\
     m06m08E &  0.13  & 0.095 &  0.014 &   1.4 &   9.7 \\
     m08m08D &  0.45  & 0.124 &  0.015 &   3.6 &   30.6 \\
     m08m08E &  0.06  & 0.124 &  0.020 &   0.5 &   2.8  \\
\enddata
\end{deluxetable}

\begin{deluxetable} {lcccrr} 

\tablecaption{Lifetimes and Population Ratios of Normal Stars \label{table:lifetimesNormal}}
\tablehead{
\colhead{Star} & \colhead{MS} & \colhead{HB} & \colhead{AGB} & \colhead{MS/HB} & \colhead{MS/AGB} \\
 & \colhead{Gyr} & \colhead{Gyr} & \colhead {Gyr} & & \\
}
\startdata
     0.78 &  15.1  & 0.097 &  0.016 &   156.5 &   924.5 \\
     0.98 &  6.50  & 0.091 &  0.016 &   71.6  &   416.7 \\
     1.18 &  3.19  & 0.092 &  0.015 &   34.6  &   212.7 \\
     1.38 &  1.81  & 0.094 &  0.014 &   19.3  &   128.5 \\
     1.58 &  1.17  & 0.095 &  0.013 &   12.3  &    90.0 \\
\enddata
\end{deluxetable}
 
The main sequence lifetimes of collision products are shorter than
normal stars (as seen in previous studies \citet{1997ApJ...487..290S,
  2001ApJ...548..323S}), with the lower mass parents showing a smaller
effect. This has already been pointed out in the evolutionary tracks,
where the main sequence extent is shorter for collision products with
more evolved parents.

For all our stars, the horizontal branch lifetime was very uniform --
approximately $1 \times 10^8$ years, regardless of parent star masses
or age at the time of the collision. However, the main sequence
lifetimes of these stars can change by 3 orders of magnitude, so the
predicted number of HB stars for every BSS depends strongly on the
kinds of BSS that are seen in clusters. All our collisions that
resulted from two 0.4 \msun~ stars colliding will not show up as BSS,
so they can be excluded from the comparison with data. Their progeny,
as well, will be indistinguishable from the normal HB stars in the
cluster.

The asymptotic giant branch lifetime was equally uniform, at about
$1.5 \times 10^7$ years. The normal stars also have the same
horizontal branch and asymptotic giant branch lifetimes as each other
and as the collision products. Post-main sequence evolution of
low-mass stars is extremely robust and largely independent of initial
mass, structure and composition profiles. 

The observed ratios of E-BSS to BSS stars are usually less than $\sim
10$.  None of the normal star models have lifetime ratios that are in
this range, although the highest mass star (m=1.58 \msun) comes
closest. We can exclude the m04m04 collision models as well, both
because their main sequence to horizontal branch lifetime ratio is so
high, and because we could not distinguish either the main sequence or
horizontal branch stars as blue stragglers observationally. Most
observers restrict themselves to `bright' blue stragglers when
constructing this ratio, which means stars more than $\sim 0.6$
magnitudes brighter than the main sequence turnoff, or a difference of
0.24 in $\log(L/L_{\odot})$. The m04m06 collision products can fall
within this range, and so it is not clear if we should be considering
them in our comparison with the observations. If we do, then the
average ratio of main sequence to horizontal branch lifetimes is 17.7;
if we only include the m04m08, m06m06, m06m08, and m08m08 collisions,
our average better matches the observations at 9.3. We note that the
observational data for M80, which is of higher quality than for the
other clusters, gives a BSS to E-BSS ratio of 16
\citep{1999ApJ...522..983F}. Because of the higher quality, blue
straggler selection was not an issue, and so the quoted ratio is for
all blue stragglers, not only the bright ones. Therefore, our
calculated average of 17.7 from all our models is in remarkably good
agreement.

From these lifetime ratios, we can also conclude that blue stragglers
must have short main sequence lifetimes. If they behaved like normal
low-mass globular cluster stars, then we would expect on the order of
100 main sequence stars for every evolved, horizontal branch phase
star. Since we see $\sim 10 - 20$ BSS for every E-BSS, our
calculations point to higher mass collision parents and a late time of
collision. Blue stragglers should have lifetimes of $\sim 1-2$
Gyr to match the observed population ratios in M3, M13, and M80, in
accordance with our models.

\begin{figure}
\includegraphics[width=\columnwidth]{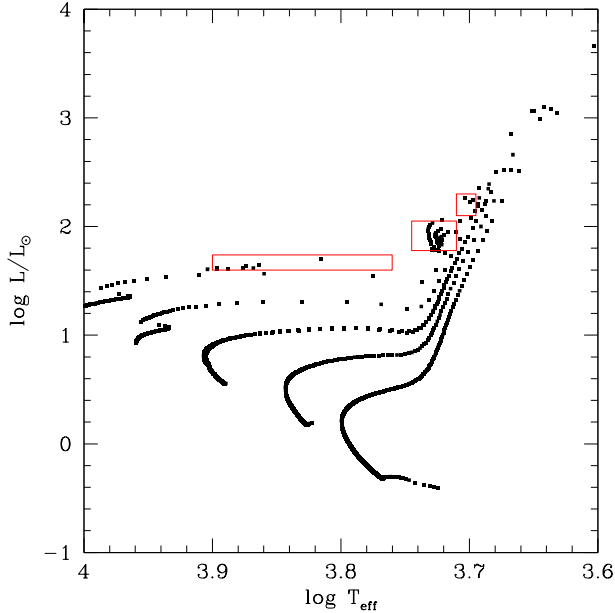}
\caption{Evolutionary tracks for all collision products for collisions
which occurred at time D, 10 Gyr after the cluster was formed. The
points are equally spaced at $10^7$ year intervals. The boxes
outline the HB, E-BSS and AGB regions of the CMD.}
\label{fig:equal1}
\end{figure}

\begin{figure}
\includegraphics[width=\columnwidth]{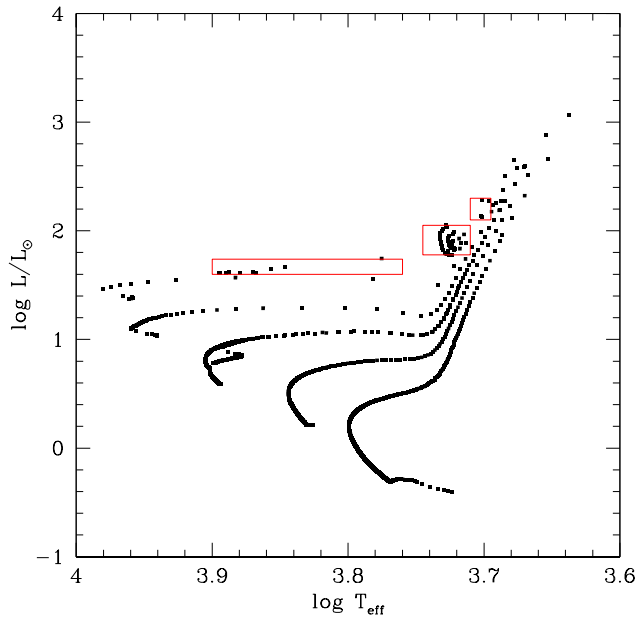}
\caption{Same as figure \ref{fig:equal1} but for collisions which
  occurred at time E, 13.7 Gyr after the cluster was formed. The boxes
outline the HB, E-BSS and AGB regions of the CMD.}
\label{fig:equal2}
\end{figure}

In figures \ref{fig:equal1} and \ref{fig:equal2}, we show the
evolutionary tracks for the collisions which occurred at times D and
E, respectively. In these figures, the points are equally spaced at
$10^7$ year intervals. By plotting the tracks in this way, it is
easier to see where the stars would lie in a cluster's CMD. The main
sequence is very well populated, since it is a long-lived phase. The
RGB coverage gets sparser towards the top, as the stars' evolution
speeds up. The core-helium burning phase is reasonably long lived, and
then there are a scattering of points on the AGB, as expected. We see
very few significant differences between these two CMDs. If we assume
that blue stragglers are formed at a constant rate in clusters (at
least over the past $\sim$3 Gyr), then the CMDs of real clusters
should look something like these diagrams, and we will not be able to
tell much about the formation rate except by looking at the bright end
of the main sequence. 

Clearly we have too many points on the main sequence -- real clusters
have a few tens to a few hundred blue stragglers per cluster, rather
than the thousands of points seen in these figures. We are also
showing an artificial concentrate of stars along the tracks we have
chosen, when of course blue stragglers could be formed with any mass
between 0.8 and 1.6 \msun~ in this scenario. However, it is
interesting to note that the E-BSS stars all clump in the same region,
regardless of mass. Therefore, the presence of E-BSS stars in a
cluster will give us very little information about the masses of their
progenitors on the main sequence, other than the fact that they are
slightly more massive than the normal turnoff (greater than 1 \msun~
compared to 0.8 \msun~ in this work).

The effect of using the \citet[][hereafter VW93]{vw93} mass loss 
during the AGB instead of Reimer's results in  $\sim 10-20$\% 
increases in the AGB lifetimes for some of the lower-mass collision 
models (e.g., m04m04A--D). 
This difference disappears in the more massive models, where the 
m06m06 and m08m08 models have essentially the same AGB lifetime 
regardless of the adopted mass-loss prescription. 
The small differences
in AGB lifetimes support our conclusions that the post-main sequence 
evolution of E-BSS is fairly robust, with the numbers of stars on
the AGB not overly dependent on input parameters such as mass loss.

The final core masses do depend on the choice of mass-loss law used
during the AGB, with VW93 generally producing larger final core masses
compared to the Reimer's formula.  For example, the m04m04 VW93 models
had final core masses $\approx 0.625$\msun compared to $0.530$\msun
for the Reimers models.  The difference was largest in the m04m06 set,
where the VW93 models resulted in core masses of 0.72\msun compared
to 0.58\msun using Reimers.  For the m06m08 and m08m08 sets of
models the difference had essentially disappeared, with both mass-loss
prescriptions giving similar core masses of 0.68\msun, with the
Reimers producing slightly larger final masses. This difference can be
explained by noting that the VW93 formula gives low mass-loss rates
($\sim 10^{-7}$\msun per year) until the pulsation period reaches
500~days, after which a superwind begins.  In the models (e.g.,
m04m06) with the largest core masses, the superwind phase is never
reached. The core keeps growing until the envelope mass is lost, as in
all stars, but the timescale for envelope loss is much longer without
the superwind.
The final core masses can be compared to WD mass distributions for
clusters \citep[e.g., NGC 6397][]{2007ApJ...671..380H}. According to Fig.~25 from
\citet{2007ApJ...671..380H}, core masses as high as 0.72\msun for $\sim$ 1\msun
progenitors are possibly not realistic, although we note that the
E-BSS will evolve somewhat differently to normal single stars.  Deep
observations of WD luminosity functions in globular clusters may be
able to tell us something about the early blue straggler population in
clusters. However, disentangling evolved blue stragglers from normal
white dwarfs will be a challenge.

\section{SUMMARY AND DISCUSSION} \label{discussion}

We calculated structural models of products of collisions between main
sequence stars, and evolved those models along the main sequence, red
giant branch, through the helium flash, on to the horizontal branch,
and through the asymptotic giant branch phase. 

We found that the models of evolved blue stragglers lie in the correct
place in the CMD to account for the observed E-BSS stars identified by
a number of groups over the past few decades. The lifetimes of these
stars are consistent with the number of observed E-BSS stars compared
to the number of blue straggler stars. We conclude that the stars
identified as E-BSS by various authors are indeed the progeny of blue
stragglers. 

The ages of low-mass stars on the horizontal branch and asymptotic
giant branch are independent of mass, collision history or lack
thereof, and initial composition profiles. The main sequence
lifetimes, on the other hand, are strongly dependent on these
quantities. Therefore, the ratio of evolved to main sequence stars
strongly constrains the main sequence lifetimes. Since we see
approximately 10 blue stragglers for every E-BSS star, rather than the
$\sim hundred$ that are predicted from normal star models, we conclude
that the main sequence lifetimes of blue stragglers are quite short,
much shorter than their normal counterparts. This result has
implications for the formation mechanisms of blue stragglers. For
example, blue stragglers cannot be fully mixed on formation (which
extends their main sequence lifetimes), and in fact must have very
truncated lifetimes. This also has implications for determinations of
formation rates. If blue stragglers are a relatively short-lived
phase, they must be formed at a high rate. The observational
identification of the E-BSS stars as high-mass, core-helium burning
stars must be robust in order to make these inferences about blue
straggler formation. Any observational work that can provide direct
evidence about the properties of the E-BSS stars will be extremely
helpful in this question. Also, any observations which can identify
E-BSS on the AGB could potentially provide more constraints on the
evolutionary status of blue stragglers, based on the models presented
here.

Our results are dependent on our determinations of the main sequence
and horizontal branch ages of all our models. While the main sequence
ages are very robustly determined, the horizontal branch ages could
depend on our treatment of the helium flash, or of the border of the
convective core. However, any of these effects must be relatively
small. First, our HB ages of normal stars are consistent with those
found in the literature for low-metallicity stars. Secondly, the HB
ages of both the normal and collisional models are very similar. Any
error introduced by an incomplete treatment of the detailed stellar
physics must be both small and constant across our models.

In this paper, we assumed that all blue stragglers were formed through
direct stellar collisions. The alternative formation mechanism, that
blue stragglers were formed through binary coalescence, may produce
stars with different structures and chemical compositions. It is
certainly expected that the main sequence evolution of a binary merger
product will be different than that of a collision product. Post-main
sequence evolution, however, is quite robust to disturbances early on
the main sequence. Even for our collision products with different
initial amounts of helium, or different convective structures, the
post-main sequence evolution is almost identical. There are slight
differences towards the end of the AGB phase and into the thermal
pulses, but between the turnoff and the tip of the AGB, the tracks,
and the timescales, are the same. This suggests that unless binary
mergers result in a star with a significantly different structure or
composition, the post-main sequence evolution will be the same as
presented here. The differences in structure or composition will need
to be quite extreme -- for example, if a binary merger product loses a
lot more mass for the same parent stars, significantly increasing the
total helium content of the star. There are almost no models for blue
stragglers formed during mass transfer. The only paper to address this
issue in detail so far is the work of \citet{2006A&A...455..247T}, who
look at mass transfer in systems appropriate to forming blue
stragglers in the open cluster M67. While they do not compare their
evolutionary tracks to normal stars, the tracks of the blue stragglers
look quite normal after mass transfer ceases. Therefore, we expect
that blue stragglers formed through binary coalescence will have
post-main sequence evolution that is very similar to that shown in
this paper, and will also populate the same position in the CMD. 

On the other hand, the main sequence lifetimes of binary coalescence
blue stragglers are uncertain. The total amount of helium in a binary
coalescence blue straggler should be the same as that formed from a
collision of the same parent stars. However, it could be distributed
differently in the star, and could affect the main sequence evolution.
If the lifetimes span the same range as the collision products in this
paper, the number of E-BSS stars in the CMD will probably not
distinguish between the different formation mechanisms, despite our
greatest hopes. Further models of binary coalescence and the
subsequent evolution of those blue stragglers will answer that question.

\acknowledgements 
This research has been supported by NSERC.
AIK acknowledges support from the Australian Research 
Council's Discovery Projects funding scheme (project number DP0664105).

\end{document}